\documentclass[prb,twocolumn]{revtex4}

\usepackage{graphicx}% Include figure files
\usepackage{dcolumn}% Align table columns on decimal point
\usepackage{bm}% bold math
\usepackage{indentfirst}

\begin{document}

\title{Single wall carbon nanotube double quantum dot}

\author{H. I. J\o rgensen}
\email{hij@fys.ku.dk}
\affiliation{Nano-Science Center, Niels Bohr
Institute, University of Copenhagen, Universitetsparken 5,
DK-2100~Copenhagen \O , Denmark}
\author{K. Grove-Rasmussen}
\affiliation{Nano-Science Center, Niels Bohr Institute, University
of Copenhagen, Universitetsparken 5, DK-2100~Copenhagen \O ,
Denmark}
\author{J. R. Hauptmann}
\affiliation{Nano-Science Center, Niels Bohr Institute, University
of Copenhagen, Universitetsparken 5, DK-2100~Copenhagen \O ,
Denmark}
\author{P. E. Lindelof}
\affiliation{Nano-Science Center, Niels Bohr Institute, University
of Copenhagen, Universitetsparken 5, DK-2100~Copenhagen \O ,
Denmark}

\date{\today}
\begin{abstract}
We report on two top-gate defined, coupled quantum dots in a
semiconducting single wall carbon nanotube, constituting a tunable
double quantum dot system. The single wall carbon nanotubes are
contacted by titanium electrodes, and gated by three narrow top-gate
electrodes as well as a back-gate. We show that a bias spectroscopy
plot on just one of the two quantum dots can be used to extract the
addition energy of both quantum dots. Furthermore, honeycomb charge
stability diagrams are analyzed by an electrostatic capacitor model
that includes cross capacitances, and we extract the coupling energy
of the double quantum dot.
\end{abstract}

\maketitle

Electronic transport in single quantum dots (QDs) defined in single
wall carbon nanotubes (SWCNTs) has been studied intensively over the
last decade.\cite{tans,nygaard,liang} These devices are typically
made by placing metal electrodes directly on top of a SWCNT
resulting in tunnel barriers at each SWCNT-metal interface, and
gated by using the substrate as one global gate. Recent studies have
shown that it is possible to locally gate and locally deplete a
small segment of a SWCNT.\cite{biercuk,tombler} By placing several
such local gates on top of a SWCNT, a double quantum dot (DQD) with
tunable inter-dot coupling can be
made.\cite{biercuk2,mason,sapmaz,graber,graber2} A DQD is a
desirable system since it can be used in the field of quantum
computation as e.g.\ a single charge qubit or two interacting spin
qubits.\cite{loss} The advantage of making DQDs in SWCNTs instead of
other material systems such as GaAs/AlGaAs is that SWCNTs are
thought to have a longer spin decoherence time. An important source
of decoherence is the hyperfine coupling between the electron in the
QD and the nuclei. $^{12}$C is the abundant isotope in natural
carbon and has no net nuclear spin. The hyperfine coupling will
therefore be highly reduced in SWCNTs.

In this Letter we present a fabrication scheme to contact and place
three narrow local gates on top of a SWCNT. We show that a device
fabricated by this method can be used to define two coupled QDs in
series. The addition energies of both QDs are estimated from a low
temperature bias spectroscopy plot on just one of the QDs. These
addition energies are then used together with a honeycomb charge
stability diagram to estimate the electrostatic coupling energy of
the DQD.

The devices are made on a highly doped silicon substrate capped by a
0.5\,$\mu$m thermally oxidized SiO$_2$ layer, and we use the
substrate as a back-gate to tune the global potential of the SWCNT.
\begin{figure*}
\begin{center}
\includegraphics[width=0.9\textwidth]{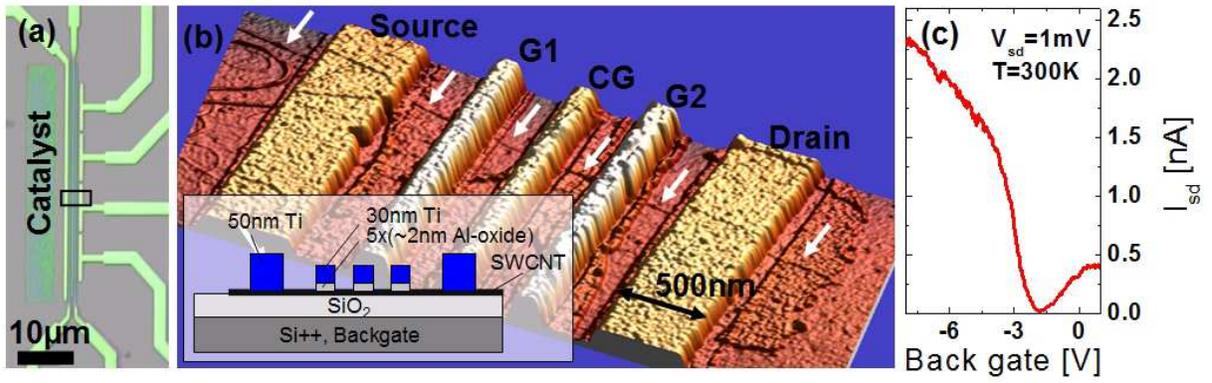}
\end{center}
\caption{(Color) (a) Optical image of 4 potential devices consisting
of one common source electrode, three common top-gate electrodes,
and four individual drain electrodes. On the left hand side of the
source electrode an island of catalyst material is positioned from
where the carbon nanotubes grow. (b) Atomic force microscope
micrograph of the region indicated by the black rectangle in (a). To
the left (close to the catalyst island) several tubes can be seen,
but only one tube has grown several $\mu$m away from the island
(indicated with white arrows). Source and drain electrodes
consisting of 50\,nm titanium, and three top-gate electrodes
consisting of five 2\,nm layers of air-oxidized aluminum and 30\,nm
titanium, are positioned directly on top of the tube. Some resist
residue can be seen around some of the leads. Insert: Schematic side
view of the device. (c) Current through the device as function of
voltage applied to the back-gate at room temperature, and with 1\,mV
source-drain voltage. \label{fig1}}
\end{figure*}
A set of alignment marks are made by electron beam lithography
(EBL), which are used to accurately position the following three
steps of EBL. First, islands of catalyst material consisting of a
suspension of aluminum oxide nanoparticles in methanol with
dissolved iron nitrate and molybdenum acetate are placed at specific
positions, see Fig.\ \ref{fig1}(a). For easy liftoff and an even
distribution of the catalyst we use a thick double layered resist
(9\% copolymer, and 4\% PMMA) and spin on the liquid catalyst at
1000\,rpm for 150\,s. The SWCNTs are then grown by chemical vapor
deposition from the catalyst islands in a ceramic tube furnace at
$\sim 900^{o}$C with a controlled flow of gasses, Ar: 1.1\,L/min,
H$_{2}$: 0.1\,L/min, and CH$_{4}$: 0.5\,L/min.\cite{kong,kgrhij}
Typically only a few or one SWCNT will grow several $\mu$m away from
the island, see Fig.\ \ref{fig1}(b). The alignment marks are
secondly used to position source and drain electrodes consisting of
50\,nm titanium with a separation of 1.8\,$\mu$m. Since the SWCNTs
tend to bundle together into ropes within a distance of about $\sim
1$\,$\mu$m from the island, the electrode nearest to the island are
positioned $\sim 2$\,$\mu$m from island, thus favoring contact to
long straight SWCNTs. In about 30\% of our devices only one tube is
contacted. Third, three narrow gate electrodes are defined by EBL
using a thin double layer resist (6\% copolymer, and 2\% PMMA) and
positioned between the source and drain electrodes, by use of the
alignment marks. The gates consist of five evaporations of aluminum
each 2\,nm thick and oxidized in air for about 1\,min, and a top
layer of titanium. We contact the EBL-structures with a final step
of optical lithography to be able to bond the device onto a
chip-carrier.

The fabrication scheme presented here has good possibilities to be
scaled up to produce several devices in each batch. In Fig.\
\ref{fig1}(a) we show a pattern with four potential devices. Several
of these patterns could easily be made in each batch, where we
currently make just two. An atomic force microscope (AFM) micrograph
of a finished device is shown in Fig.\ \ref{fig1}(b) where only one
SWCNT is contacted. The three gates are named G1, CG (center gate),
and G2 starting from the source electrode. We apply source-drain
voltage ($V_{sd}$) to the source electrode and keep the drain
electrode at ground. The nanotube in the device that we present
measurements on in this Letter has a height (diameter) measured with
an AFM of about $\sim1$\,nm. It shows an ambipolar characteristic at
room temperature as seen in Fig.\ \ref{fig1}(c), which suggests that
it is a small band gap semiconducting SWCNT. We can thus use the
back-gate to tune the global potential of the device from electron
to hole transport. In the rest of the letter the measurements are
made through the valence band with a back-gate voltage of
$V_{BG}=-6$\,V, to ensure that transport is governed by holes.

\begin{figure}
\begin{center}
\includegraphics[width=0.48\textwidth]{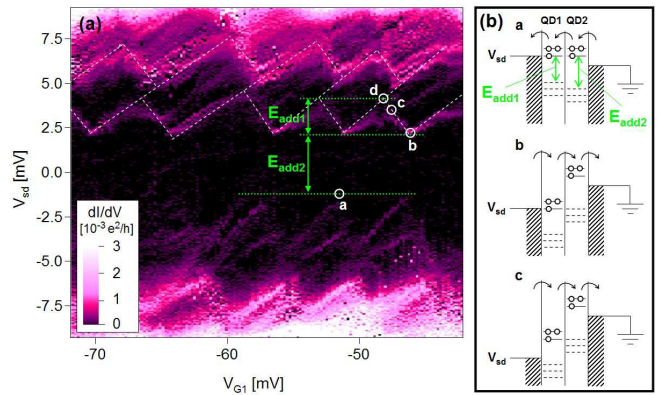}
\end{center}
\caption{(Color) (a) Bias spectroscopy plot of differential
conductance (dI/dV) versus source-drain voltage and voltage applied
to G1, with $V_{CG}=0$\,V, $V_{G2}=1.1$\,V, and $V_{BG}=-6$\,V at
300\,mK. The white dashed lines are guidelines to the eye,
indicating charge degeneracy lines. The addition energies of each
quantum dot are indicated with green arrows. (b) Schematic figures
of the hole transport through the double quantum dot at positions
indicated with letters in (a). Solid and dashed lines are filled and
empty hole states, respectively. The coupling energy is here
neglected since it is much smaller than the addition energies (see
below). \label{fig2}}
\end{figure}

Figure\,\ref{fig2}(a) shows a bias spectroscopy plot at 300\,mK of
the differential conductance versus $V_{sd}$ and voltage applied to
G1 ($V_{G1}$), with CG and G2 kept constant at $V_{CG}=0$\,V, and
$V_{G2}=1.1$\,V, respectively. That is, in Fig.\,\ref{fig2}(a) QD1
is probed by the source electrode from the left hand side, and a
discreet energy level of QD2 from the right hand side.
Around zero bias the device does not conduct, and the onset of
conductance is asymmetric around zero bias. The onsets of
conductance at point \textbf{a} and \textbf{b} in Fig.\
\ref{fig2}(a) are positioned at $V_{sd} \sim -1.5$\,mV, and $V_{sd}
\sim 2.1$\,mV, respectively. The conductance gap is constant in the
bias spectrum in Fig.\ \ref{fig2}(a), and also constant in the whole
gate range that we measured ($V_{G1}=\pm 100$\,mV). This gap in
conductance is due to the DQD nature of the device, where the first
QD (QD1) is tuned by G1, and the second QD (QD2) is tuned by G2.
Both QDs have Coulomb blockade (CB) oscillations and since QD2 is in
CB for the chosen gate voltage on G2, transport is blocked whenever
the bias is smaller than the addition energy of QD2. Since QD1 is
probed from the right hand side by energy levels from QD2 and
because the chemical potential of the drain lead is aligned
asymmetrically between two successive energy levels of QD2, the
conductance gap is asymmetric around zero bias (see Fig.\
\ref{fig2}(b)). At point \textbf{a} the energy levels of the two QDs
are aligned with the chemical potential of the source lead, and we
have hole transport from drain to source. From point \textbf{a} to
point \textbf{b} the energy level of QD1 and the chemical potential
of the source lead are kept aligned and shifted together, while QD2
is kept constant in CB, i.e., no sequential tunneling is possible.
At point \textbf{b} the chemical potential of the source lead and
the energy level of QD1 becomes aligned with the next energy level
of QD2, which gives hole transport from source to drain. The
conductance gap is therefore a measure of the addition energy of QD2
($E_{add2}$).

Above and below the conductance gap structures similar to CB
diamonds for a single QD are observed. These structures are due to
CB in QD1 and illustrated from point \textbf{b} to point \textbf{d}
through point \textbf{c} in Fig.\ \ref{fig2}(a). Along the line from
point \textbf{b} to point \textbf{d} the ground level in each QD are
kept aligned, while the chemical potential of the source electrode
is shifted downwards to align with the next energy level of QD1.
Because of the capacitive coupling between source and QD1 a negative
compensating gate voltage on G1 is needed to keep the ground levels
in QD1 and QD2 aligned. The distance from point \textbf{b} to point
\textbf{d} in source-drain voltage is therefore a measure of the
addition energy of QD1 ($E_{add1}$). Since no odd/even or
four-period structures originating from the level spacings in the
two dots is observed in either bias spectroscopy plots or honeycomb
charge stability diagrams (see below), we estimate the level
spacings to be much smaller than the charging energies. The two
methods to read-off the addition energy of QD1 and QD2 gives on
average $E_{add1} \sim 2.2$\,meV, and $E_{add2} \sim 3.6$\,meV. At
higher bias (above the level of point \textbf{d}) more structures
are observed. A thorough explanation of these structures is beyond
the scope of this letter but an interesting subject for further
study.

\begin{figure}
\begin{center}
\includegraphics[width=0.48\textwidth]{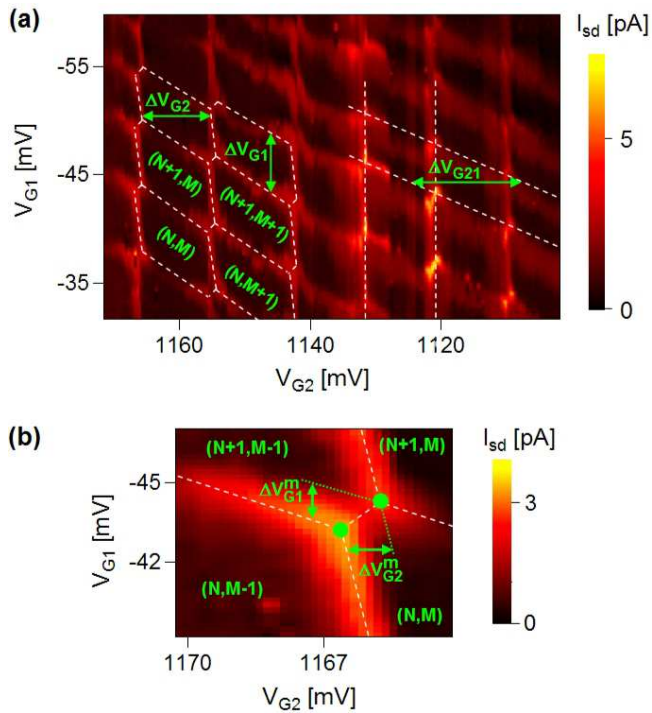}
\end{center}
\caption{(Color) Charge stability diagrams at 300\,mK of the
measured current as function of $V_{G1}$, and $V_{G2}$, with
$V_{CG}=0$\,V, $V_{BG}=-6$\,V, and $V_{sd}=2$\,mV. (a) Honeycomb
pattern with relative number of holes in each QD indicated with
green numbers. The white dashed lines are guidelines to the eye. (b)
Close-up of one set of triple points at the position indicated with
the relative hole numbers. \label{fig3}}
\end{figure}

Figure\,\ref{fig3} shows a charge stability diagram of current
through the DQD as function $V_{G1}$, and $V_{G2}$.
Honeycomb structures can be identified throughout the plot which is
a clear sign of a DQD with inter-dot coupling. Within each honeycomb
structure the number of holes in each QD is constant, as indicated
with relative hole numbers (N,M) in Fig.\ \ref{fig3}(a) and (b). At
the corners of these honeycombs so-called triple points are located,
where three charge states are degenerate, e.g., (N,M), (N+1,M), and
(N,M+1). At these triple points an increase in current is observed
consistent with sequential tunneling becoming possible via the three
degenerate charge states. Furthermore, the overall slope of the
honeycombs as illustrated with white dashed lines in the right side
of Fig.\ \ref{fig3}(a) can be used to estimate cross capacitances.
When G2 is decreased by $\Delta V_{G21}$ (indicated in Fig.\
\ref{fig3}(a)) one hole is added to QD1, i.e., a cross capacitance
from G2 to QD1 exists. Since the vertical distance ($\Delta
V_{G12}$) between the two almost vertical lines to the right in
Fig.\ \ref{fig3}(a) tends to infinity, there are almost zero cross
capacitance from G1 to QD2.

The observed splitting of adjacent triple points, as shown in Fig.\
\ref{fig3}(b) is due to coupling between the QDs. The electrostatic
coupling energy ($E_{Cm}$) is defined as the change in potential
energy of QD1 when a hole is added to QD2, or vice versa. We have
extended the electrostatic capacitor model in Ref.\,[14] to include
cross capacitances. We find that the electrostatic coupling energy
can be given in terms of quantities directly observable in a bias
spectroscopy plot and in a honeycomb charge stability diagram:
\begin{equation}\label{eqa}
E_{Cm} = E_{add1(2)} \cdot \frac{\Delta V_{G1(2)}^{m}}{\Delta
V_{G1(2)}} \cdot \frac{\Delta V_{G12(21)}}{\Delta V_{G12(21)}-\Delta
V_{G1(2)}^{m}}
\end{equation}
where $\Delta V_{G1(2)}$, and $\Delta V_{G1(2)}^{m}$, is the size of
the honeycombs and the splitting of the triple points, as
illustrated in Fig.\ \ref{fig3}(a) and (b), respectively. The last
term in Eq.\ (\ref{eqa}) accounts for the cross capacitances and
goes to unity when there are no cross capacitances, i.e., $\Delta
V_{G12(21)}$ goes to infinity. An average estimate of $\Delta
V_{G1(2)}$ from all the honeycombs seen in Fig.\ \ref{fig3}(a) gives
$\Delta V_{G1(2)} \sim 6(10)$\,mV. The estimated values of $\Delta
V_{G1(2)}^{m}$ and $\Delta V_{G12(21)}$ is; $\Delta V_{G1(2)}^{m}
\sim 1.25(1.10)$\,mV, and $\Delta V_{G21} \sim 20mV$ ($\Delta
V_{G12}$ tends to infinity). Since the gate voltages used in the
bias spectrum in Fig.\ \ref{fig2} and the gate voltages used in the
charge stability diagram in Fig.\ \ref{fig3} are chosen to be
roughly the same, we can use the addition energies found above in
the estimation of $E_{Cm}$. From these experimental values two
consistent estimates of the electrostatic coupling energy is
obtained, $E_{Cm} \sim 0.46(0.42)$\,meV.

In conclusion we have presented a fabrication scheme that in each
batch can produce several devices for electronic transport in a
SWCNT with three narrow top-gates. We show that a device fabricated
by this method can be used to define two coupled QDs in series. From
a bias spectroscopy plot of just one of the QDs the addition
energies of both QDs are extracted. Furthermore, an electrostatic
capacitor model that includes an observed cross capacitance is used
on a honeycomb charge stability diagram to extract two consistent
estimates of the coupling energy.

We wish to acknowledge the support of the EU-STREP Ultra-1D program,
and the Nano-Science Center, University of Copenhagen, Denmark.

\end{document}